\documentclass[aps,reprint,showkeys,noshowpacs,pre]{revtex4-1}
\usepackage[utf8x]{inputenc}
\usepackage{amsmath,graphicx,color}
\definecolor{myred}{rgb}{1,0.2,0.2}
\definecolor{mygreen}{rgb}{0,0.5,0}
\usepackage[colorlinks=true, urlcolor=blue, citecolor=mygreen]{hyperref}
\newcommand{\avg}[1]{\langle #1\rangle}
\newcommand{\eref}[1]{Eq.~(\ref{#1})}
\newcommand{\vek}[1]{\boldsymbol{#1}}
\newcommand{\dd}{\mathrm{d}}
\newcommand{\ee}{\mathrm{e}}
\begin{document}
\title{Statistical validation of high-dimensional models of growing networks}
\author{Matúš Medo}
\affiliation{Physics Department, Chemin du Musée 3, University of Fribourg, 1700 Fribourg, Switzerland}

\begin{abstract}
The abundance of models of complex networks and the current insufficient validation standards make it difficult to judge which models are strongly supported by data and which are not. We focus here on likelihood maximization methods for models of growing networks with many parameters and compare their performance on artificial and real datasets. While high dimensionality of the parameter space harms the performance of direct likelihood maximization on artificial data, this can be improved by introducing a suitable penalization term. Likelihood maximization on real data shows that the presented approach is able to discriminate among available network models. To make large-scale datasets accessible to this kind of analysis, we propose a subset sampling technique and show that it yields substantial model evidence in a fraction of time necessary for the analysis of the complete data.
\end{abstract}

% PACS codes: 89.75.Hc, 64.60.aq, 87.23.Ge, 02.50.-r

\maketitle

\section{Introduction}
Complex networks are used to represent and analyze a wide range of systems~\cite{Dorogovtsev2002,Boccaletti2006,Newman2010}. Models of complex networks usually aim for simplicity and attempt to keep the number of parameters as low as possible. However, real data is more complex than any simple model which makes it difficult to draw clear links between data and models. To capture the increasingly available massive real data~\cite{Gonzales2007}, we need high-dimensional models where the number of parameters grows with the number of nodes. An example of such a model is the latent space model~\cite{Hoff02} where nodes are assigned independent and identically distributed vectors and the probability of a link connecting two nodes depends only on the distance of their vectors.

While there are plenty of simple (and not so simple) network models, little is known as to which of them are really supported by data. While calibration of complex network models often uses standard statistical techniques, their validation is typically based on comparing their aggregate features (such as the degree distribution or clustering coefficient---see~\cite{Costa2007,Kolaczsyk2009} for detailed accounts on network measurements) with what is seen in real networks (see~\cite{Papa2012,Li2013} for recent examples of this approach). The focus on aggregate quantities naturally reduces the discriminative power of model validation which is often further harmed by the use of inappropriate statistical methods~\cite{Stumpf12}. As a result, we still lack knowledge of what is to date the best model explaining the growth of the scientific citation network, for example.

We argue that network models need to be evaluated by robust statistical methods~\cite{Bos07,Freedman}, especially by those that are suited to high-dimensional models~\cite{Buehlmann2011}. This is exemplified in~\cite{Leskovec08} where various low-dimensional microscopic mechanisms for evolution of social networks are compared on the basis of their likelihood of generating the observed data. Prohibitive computational complexity of maximum likelihood estimation is often quoted as a reason for its limited use in the study of real world complex networks~\cite{Leskovec10}. However, as we shall see here, even small subsets of data allow to discriminate between models and point clearly to those that are actually supported by the data. This, together with the ever-increasing computational power at our disposal, opens the door to the likelihood analysis of complex network models.

We analyze here a recent network growth model~\cite{Medo2011} which naturally leans itself to high-dimensional analysis. This model generalizes the classical preferential attachment (PA; often referred to as the Barabási-Albert model in the complex networks literature) \cite[Sections~7,~8]{Albert2002} by introducing node relevance which decays in time and co-determines (together with node degree) the rate at which nodes acquire new links. If either the initial relevance values or the functional form of the relevance decay are heterogeneous among the nodes, this model is able to produce various realistic degree distributions. By contrast to~\cite{Eom2011} which modifies preferential attachment by introducing an additive heterogeneous term, in~\cite{Medo2011} relevance combines with degree in a multiplicative way which means that once it reaches zero, the degree growth stops. This makes the model an apt candidate for modeling information networks where information items naturally lose their pertinence with time and the growth of their degree eventually stops. (See~\cite{Holme2012} for a review of work on temporal networks.) This model has been recently used to quantify and predict citation patterns of scientific papers~\cite{Wang2013}.

Before methods for high-dimensional parameter estimation are applied to real data, we calibrate and evaluate them on artificial data where one has full control over global network parameters (size, average degree, etc.) and true node parameter values are known. For simplicity, we limit our attention to the case where the functional form of relevance decay is the same for all nodes and only the initial relevance values differ. We present here various estimation methods and evaluate their performance. Plain maximum likelihood~\cite[Chapter~7]{Freedman} produces unsatisfactory results, especially in the case of sparse networks which are commonly seen in practice. We enhance the method by introducing an additional term which suppresses undesired correlation between node age and estimates of initial relevance. We then introduce a mean-field approach which allows us to reduce high-dimensional estimation to a low-dimensional one. Calibration and evaluation of these parameter-estimation methods is done on artificial data. Real data is then used to employ the established framework and compare the statistical evidence for several low- and high-dimensional network models on the given data. Analysis of small subsets of input data is shown to efficiently discriminative among the available models. Since this work focuses on model evaluation, estimated parameter values are thus of secondary importance to us. Necessary conditions for obtaining precise estimates and the potential risk of large errors~\cite{Owhadi2013} are therefore left for future research (see Sec.~\ref{sec:conclusions}).

\section{Model}
\label{sec:model}
The original model of preferential attachment with relevance decay (PA-RD) has been formulated for an undirected network where the initial node degree is non-zero because of links created by the node on its arrival~\cite{Medo2011}. To allow zero-degree nodes to collect links, some additive attractiveness or random node selection need to be introduced. When these two mechanisms are combined with PA-RD, the probability that a new link created at time $t$ attaches to node $i$ can be written as
\begin{equation}
\label{PARD}
P(i, t) = \lambda\,\frac{R_i(t)\big(k_i(t)+A\big)}{\sum_{j=1}^{n(t)} R_j(t)\big(k_j(t)+A\big)} +
\frac{1-\lambda}{n(t)}.
\end{equation}
Here $k_i(t)$ and $R_i(t)$ are degree and relevance of node $i$ at time $t$, respectively, $n(t)$ is the number of nodes present at time $t$, and $A$ is the additive attractiveness term. Finally, $\lambda$ is the probability that the node is chosen by the PA-RD mechanism; the node is chosen at random with the complementary probability $1-\lambda$. When $A=0$ and $\lambda=1$, a node of zero degree will never attract new links. \eref{PARD} can be used to model a monopartite network where nodes link to each other as well as a bipartite network where one set of nodes is unimportant and we can thus speak of outside links attaching to nodes. For example, one can use the model to describe the dynamics of item popularity in a user-item bipartite network representing an e-commerce system~\cite{Ming-Sheng10}.

There are now two points to make. Firstly, the model is invariant with respect to the rescaling of all relevance values, $R_i(t)\to \xi R_i(t)$. This may lead to poor convergence of numerical optimization schemes because $R_i(t)$ values can drift in accord without affecting the likelihood value. The convergence problems can be avoided by imposing an arbitrary normalization constraint on the relevance values as we do below. Secondly, $A$ and $\lambda$ act in the same direction: they introduce randomness in preferential attachment-driven network growth (in particular, as $A\to\infty$ and/or $\lambda\to0$, preferential attachment loses all influence). One can therefore expect that $A$ and $\lambda$ are difficult to be simultaneously inferred from the data. This is especially true for the original preferential attachment without decaying relevance. If node relevance decays to zero, node attraction due to $A$ eventually vanishes while the random-attachment part proportional to $\lambda$ remains---it is therefore possible, at least in principle, to distinguish between the two effects. To better focus on the high-dimensional likelihood maximization of node parameters, we assume $\lambda=1$ in all our simulations.

The PA-RD model has been solved in~\cite{Medo2011} for a case where $\lambda=1$, $A=0$, and the initial degree of all nodes equal to one. It was further assumed that $T_i:=\int_0^{\infty} R_i(t)\,\dd t$ is finite for all nodes and the distribution of $T$ values among the nodes, $\varrho(T)$, decays exponentially or faster. The probability normalization term $\sum_j R_j(t)(k_j(t)+A)$ then eventually fluctuates around a stationary value $\Omega^*$ and the expected final degree of node $i$ can be written as $\avg{k_i^F}=\exp(T_i/\Omega^*)$. It has been shown that the network's degree distribution, shaped mainly by $\varrho(T)$, can take on various forms including exponential, log-normal, and power-law.

\subsection{Description of artificial data}
\label{sec:artif_data}
\begin{figure}
\centering
\vspace*{4pt}
\includegraphics[scale = 0.34]{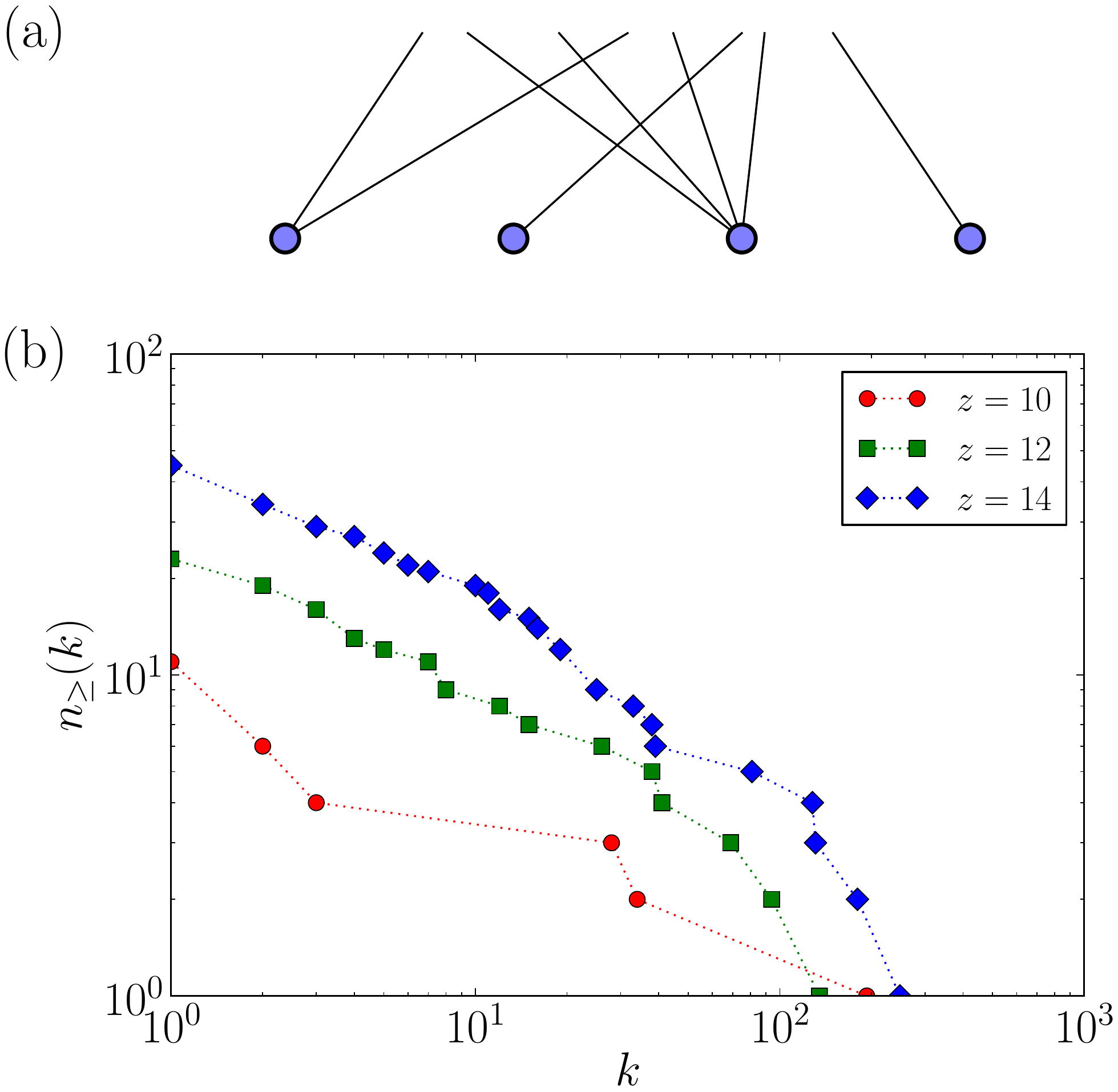}
\caption{(a) Illustration of a bipartite network where only links' target nodes are of interest. (b) Sample degree distributions of networks produced according to Sec.~\ref{sec:artif_data}.}
\label{fig:network}
\end{figure}

We begin by describing bipartite network data with temporal information. We consider a simplified bipartite case where links arrive from outside and thus only their target nodes matter---see Fig.~\ref{fig:network}a for illustration. Links are numbered with $l=1,\dots,E$ and the times at which they are introduced are $t_1\leq t_2\leq\cdots\leq t_E$. Nodes are numbered with $i=1,\dots,N$ and the times at which they are introduced are $\tau_1\leq\tau_2\leq\cdots\leq\tau_N$. At time $t$, there are $n(t)$ target nodes in the network. Degree of node $i$ at time $t_l$ when link $l$ is added is $k_i(t_l)$ and the target node of link $l$ is $n_l$. The average node degree is $z:=E/N$ (the factor of two is missing here because we consider a bipartite network where $E$ edges point to $N$ nodes of interest).

We use the PA-RD model to create artificial networks with well-defined properties. There are initially $n_I$ nodes with zero degree. After every $\Delta T$ time steps, a new node of zero degree is introduced in the network. In each time step, one new link is created and chooses its target node according to \eref{PARD}. The network growth stops once there are $E_F = \lceil zn_I / (1-z/\Delta T)\rceil$ links and $n_F = n_I + \lfloor E_F/\Delta T\rfloor$ nodes in the network. At that point, the average node degree is approximately $z$. It must hold that $z<\Delta T$; in the opposite case, the average degree $z$ cannot be achieved because new nodes dilute the network too fast. Each node has the relevance decay function $R_i(t) = I_i \exp[-(t-\tau_i)/\varTheta]$ where $\varTheta$ is the decay time scale and $I_i$ is the initial relevance of node $i$. Initial relevance values are drawn from the exponential distribution $f(I)=\ee^{-I}$. When the decay parameter $\varTheta$ is sufficiently high, this setting produces broad degree distributions~\cite{Medo2011} which are similar to distributions often seen in real information networks~\cite[Chapter~4]{Newman2010}. We use $n_I=10$, $\Delta T = 16$, $\varTheta=50$, $A=1$, and $\lambda=0$ for all artificial networks studied here; their sample degree distributions are shown in Fig.~\ref{fig:network}b.

\section{Parameter estimation methods}

\subsection{Maximum likelihood estimation}
\label{sec:MLE}
We first use the standard maximum likelihood estimation (MLE) to estimate parameters of the PA-RD model~\cite{Bos07}. A generic form of log-likelihood of realization $\mathcal{D}$ for a network growth model $\mathcal{M}$ has the form
\begin{equation}
\ln\mathcal{L}(\mathcal{D}\vert\mathcal{M}) = \sum_{l=1}^E \ln P(n_l, t_l\vert\mathcal{M}).
\end{equation}
where $P(n_l, t_l\vert\mathcal{M})$ is the probability of link $l$ arriving at node $n_l$ at time $t_l$ under model $\mathcal{M}$. It is convenient to transform this quantity into log-likelihood per link by dividing it with the number of links, $\ln\mathcal{L}(\mathcal{D}\vert\mathcal{M})/E$. For model $\mathcal{M}$ represented by its attachment probability $P(n_l, t_l\vert\mathcal{M})$ and a vector of model parameters $\vek{p}$, log-likelihood can be maximized with respect to these parameters and yields their estimates $\tilde{\vek{p}}$.

Given a network realization obtained with Eq.~(\ref{PARD}), there are several parameters to estimate: initial relevance values of all nodes, additional attractiveness term $A$, and parameters of the relevance decay function. (Note that we make the estimation task easier by assuming that the functional form of relevance decay is known.) Greedy (uphill) maximization of log-likelihood is made possible by the profile of the likelihood function which does not feature multiple local maxima in the space of initial relevance values (see Sec.~\ref{sec:maxima} for an explanation). Starting from a random initial guess, we sequentially update all model parameters by quadratic extrapolation and repeat this process until the difference between new and old estimates is less than some sufficiently small threshold (we use $10^{-3}$ here). Due to the scale-invariance of relevance values, they can be normalized after each iteration so that their average is one, which improves convergence. While each evaluation of log-likelihood is time consuming and this straightforward approach is thus computationally expensive, it is often, as we shall show, viable.

\subsection{Mean-field approximation to MLE (MF-MLE)}
\label{sec:MF-MLE}
As mentioned in Sec.~\ref{sec:model}, when the number of nodes is large and their relevance decays to zero, fluctuations of the denominator in \eref{PARD} become small and one can therefore replace it with a constant term $\Omega^*$. This mean-field approximation decouples the dynamics of nodes which then compete for new links with the external field $\Omega^*$ instead of competing with the other nodes present in the system. Eq.~(\ref{PARD}) then simplifies to
\begin{equation}
P(i, t) = \frac{R_i(t\vert\vek{\eta}_i)\big(k_i(t)+A\big)}{\Omega^*}
\end{equation}
where $\vek{\eta}$ is a vector of parameters of node $i$ and we again assume $\lambda=1$. In our case, the initial relevance value $I_i$ is the only node-specific parameter and thus $\vek{\eta}_i = (I_i)$. Since $\Omega^*$ is the same for all nodes, we can subsume it in $I_i$ due to the aforementioned scale invariance. The likelihood function for node $i$ is then constructed by evaluating all links created after this node has been introduced in the network. For link $l$, we assess whether the link points to node $i$ (then $\delta_{i, n_l} = 1$) or not (then $\delta_{i, n_l} = 0$). We get
\begin{gather}
\label{LL-mean_field}
\ln\mathcal{L}_i(\mathcal{D}\vert \vek{\eta}_i) = \notag\\
= \sum_{\genfrac{}{}{0pt}{}{l=1}{t_l\geq\tau_i}}^E
\ln\big[P(i, t_l)\delta_{i,n_l} + (1-P(i, t_l))(1-\delta_{i,n_l})\big]
\end{gather}
where we ignore links that are older than node $i$. This function can be maximized with respect to $\vek{\eta}_i$ for any given $A$. Global model parameters such as, in our case, $A$ and the time scale of relevance decay $\varTheta$ can be estimated by minimizing $\sum_i\ln\mathcal{L}_i(\mathcal{D}\vert\tilde{\vek{\eta}}_i)$ with respect to them (estimates $\tilde{\vek{\eta}}_i$ then need to be updated to reflect new values of the global parameters).

MF-MLE makes it easy to change the functional form of relevance decay $R(t)$ for any individual node and thus classify their behavior (see~\cite[Chapter~8]{Han11} for more information on classification problems). While we do not pursue this direction here, it is of particular significance to the analysis of real data where various behavioral classes of nodes are likely to coexist. Also, the vector of node parameters can be easily extended by, for example, making the decay time $\varTheta$ node-dependent, while still maintaining the low-dimensional nature of the resulting likelihood optimization.

\section{Estimation evaluation}
To evaluate various estimation methods, we assess the maximal likelihood that they are able to achieve. Parameter estimation is simplified by assuming that the functional form of relevance decay is known and only model parameters $A, \varTheta, \{I_i\}_i$ are to be estimated. Since the true parameter values are available to us, we also measure Pearson's correlation between true values $I$ and their estimates $\tilde I$, $r(I, \tilde I)$ (the higher the value, the better the estimates). In evaluating this correlation, nodes with final degree four and less are excluded because their estimates are too noisy due to the lack of data. The advantage of using Pearson's correlation to measure the accuracy of estimates lies in its invariance with respect to rescaling of $\tilde I_i$ which fits well with the scale-invariance of the PA-RD model itself. The accuracy of estimates of $A$ and $\varTheta$ is measured as well.

\begin{figure}
\centering
\includegraphics[scale = 0.34]{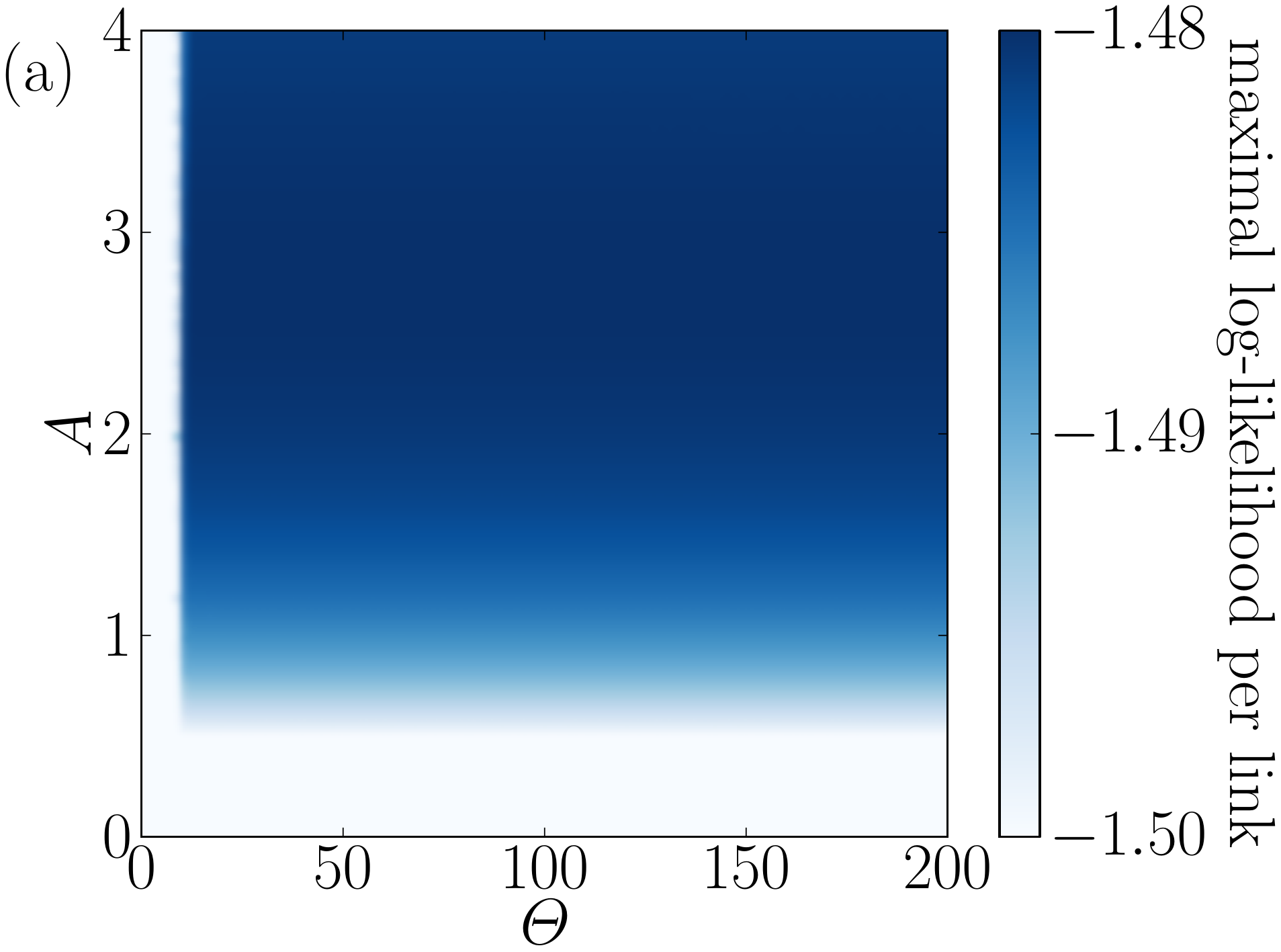}\\[4pt]
\includegraphics[scale = 0.34]{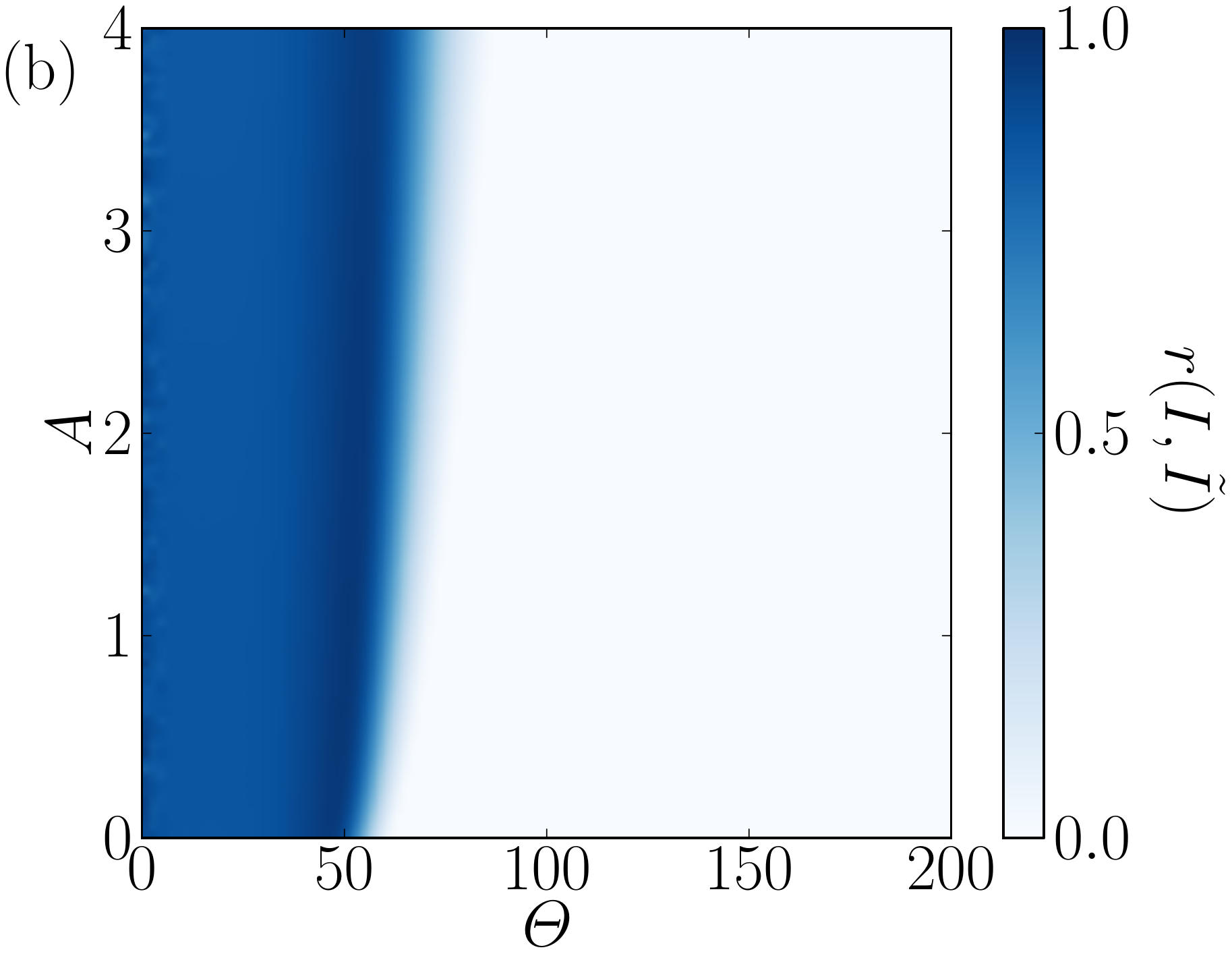}
\caption{Results of a constrained MLE procedure where $A$ and $\varTheta$ are fixed at various values and log-likelihood is thus maximized only with respect to the initial relevance of each node. These results were obtained for one realization of the artificial network model with $z=12$ which corresponds to $n_F=30$ and $E_F=480$.}
\label{fig:profile}
\end{figure}

Simulations reveal that MLE sometimes converges to estimates which are far from the true parameter values. To explain the reason for this behavior, Fig.~\ref{fig:profile} shows the results of constrained likelihood maximization where we artificially fix $A$ and $\varTheta$ at various values, many of which are far from the true values $A=1$ and $\varTheta=50$. The corresponding maximal log-likelihood values exhibit a shallow maximum in $A$ with the optimal value $2.7$ lying significantly above the true value $1$. Worse, the maximum in $\varTheta$ is non-existent: as $\varTheta$ increases, log-likelihood increases too and saturates at a value which is maintained also in the limit $\varTheta\to\infty$ (\emph{i.e.}, no relevance decay). Resulting $\tilde\varTheta$ thus depends on the initial values of model parameters and the procedure in which they are iteratively improved in the search for maximal likelihood. While Fig.~\ref{fig:profile} shows results for one network realization, the same behavior can be seen for all realizations of the input artificial network. Inspection of the initial relevance values estimated for large $\varTheta$ makes it clear that the lack of relevance decay is then compensated by later nodes being assigned higher initial relevance than earlier ones. As a result, MLE estimates then do not reflect the true initial relevance values but rather the order in which nodes are introduced in the network. This is demonstrated by the second panel of Fig.~\ref{fig:profile} where $r(I, \tilde I)$ reaches maximum for $\varTheta$ close to the true value of $50$ and then quickly drops to negative values for larger values of $\varTheta$. The negative correlation values are observed here because in this particular network realization, node arrival times are negatively correlated with their initial relevance values. The overall maximum of $r(I,\tilde I)$ lies at $A=0.94$ and $\varTheta=51$.

\begin{figure}
\centering
\includegraphics[scale = 0.4]{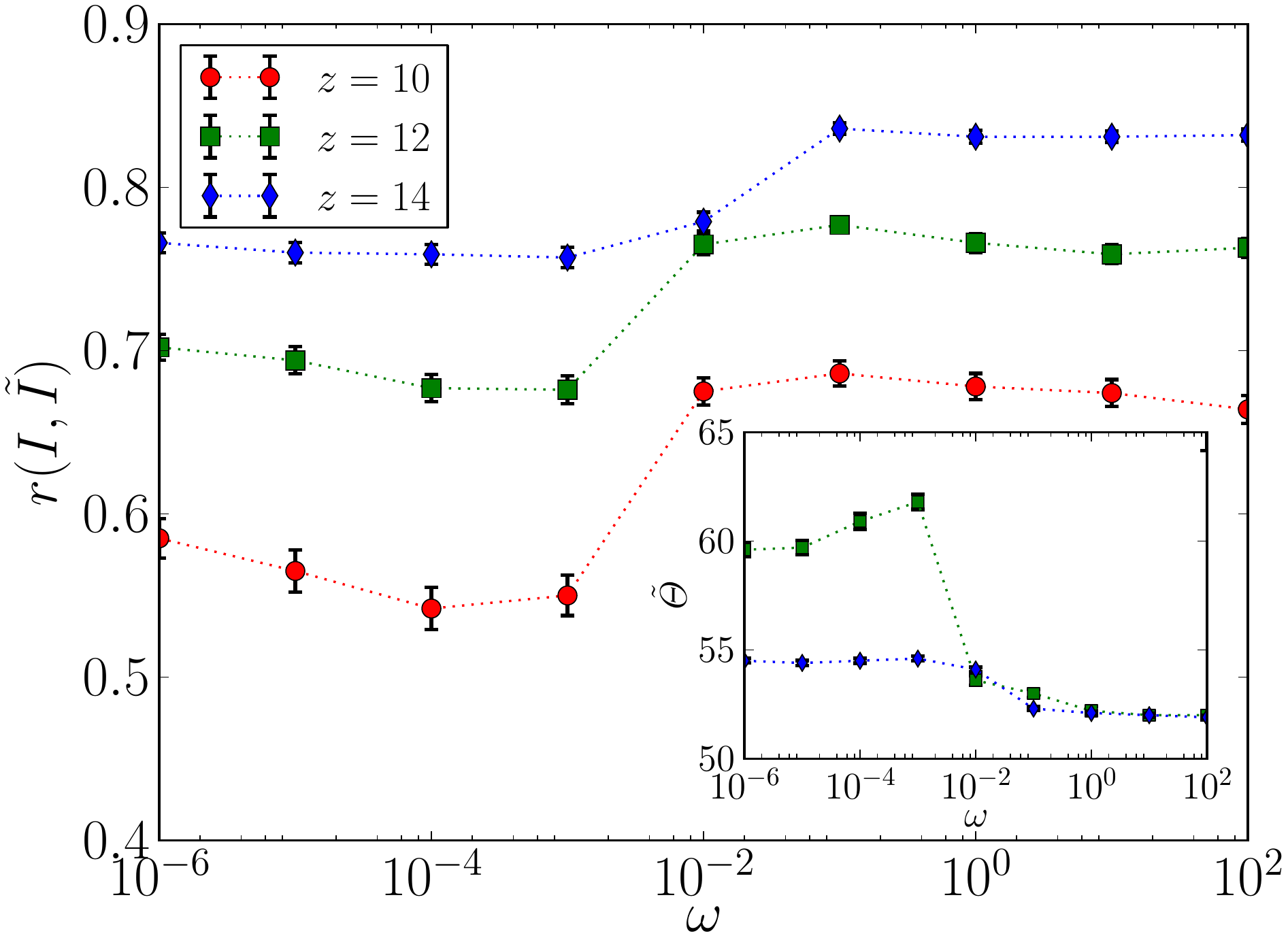}
\caption{Impact of the log-likelihood penalization term given in \eref{LLpenalized} as a function of $\omega$ for various values of $z$. Mean values and their standard errors were obtained by estimation in 1000 independent network realizations (the same applies to Tab.~\ref{tab:evaluation} and Fig.~\ref{fig:comparison}). Estimates of $\varTheta$ are large for $z=10$ and thus missing in the inset.}
\label{fig:omega}
\end{figure}

The problem of excessive estimated decay time $\varTheta$ can be solved by introducing an additional term in log-likelihood with the aim to penalize solutions with high $\varTheta$. This is similar to regularization schemes such as LASSO~\cite{Tibshirani1996} which are often used to constraint solutions in high-dimensional optimization problems~\cite{Buehlmann2011}. We choose here to maximize
\begin{equation}
\label{LLpenalized}
\frac1E\ln\mathcal{L}_i(\mathcal{D}\vert I_i, A, \varTheta)-
\omega r(\tau,\tilde I) g[r(\tau,\tilde I)]
\end{equation}
where $g(x) = x$ for $x>0$ and $0$ otherwise; the additional term penalizes positive correlation between nodes' arrival times and their estimated initial relevances. As shown in Fig.~\ref{fig:omega}, MLE estimates with the correlation term $r(\tau, \tilde I)$ are superior to the original ones over a broad range of $\omega$. The difference is particularly large for sparse networks where standard MLE strongly overestimates $\varTheta$. Note that unlike for large parts of Fig.~\ref{fig:profile}b, the average correlation value $r(I,\tilde I)$ in Fig.~\ref{fig:omega} is positive even when $\tilde\varTheta$ is large. This is because while Fig.~\ref{fig:profile} presents outcome for a single network realization, Fig.~\ref{fig:omega} averages over many of them and the resulting average correlation is positive.

\begin{table}
\centering
\begin{ruledtabular}
\begin{tabular}{rrrrr}
         method & $\ln\mathcal{L}/E$ & $\tilde A$ & $\tilde\varTheta$ & $r(I, \tilde I)$\\
\hline
MLE, $\omega=0$ & $-1.350(5)$ & $2.86(2)$ & $54.4(1)$ & $0.767(5)$\\
MLE, $\omega=1$ & $-1.350(5)$ & $2.85(2)$ & $51.8(1)$ & $0.835(3)$\\
         MF-MLE & $-1.381(5)$ & $6.09(9)$ & $65.2(2)$ & $0.722(2)$\\
\end{tabular}
\end{ruledtabular}
\caption{Estimates obtained with respective methods for $z=14$ (which corresponds to $n_F=80$, $E_F=1120$). Numbers in brackets report uncertainty of the last digit given by the standard error of the mean.}
\label{tab:evaluation}
\end{table}

\begin{figure}
\centering
\includegraphics[scale = 0.4]{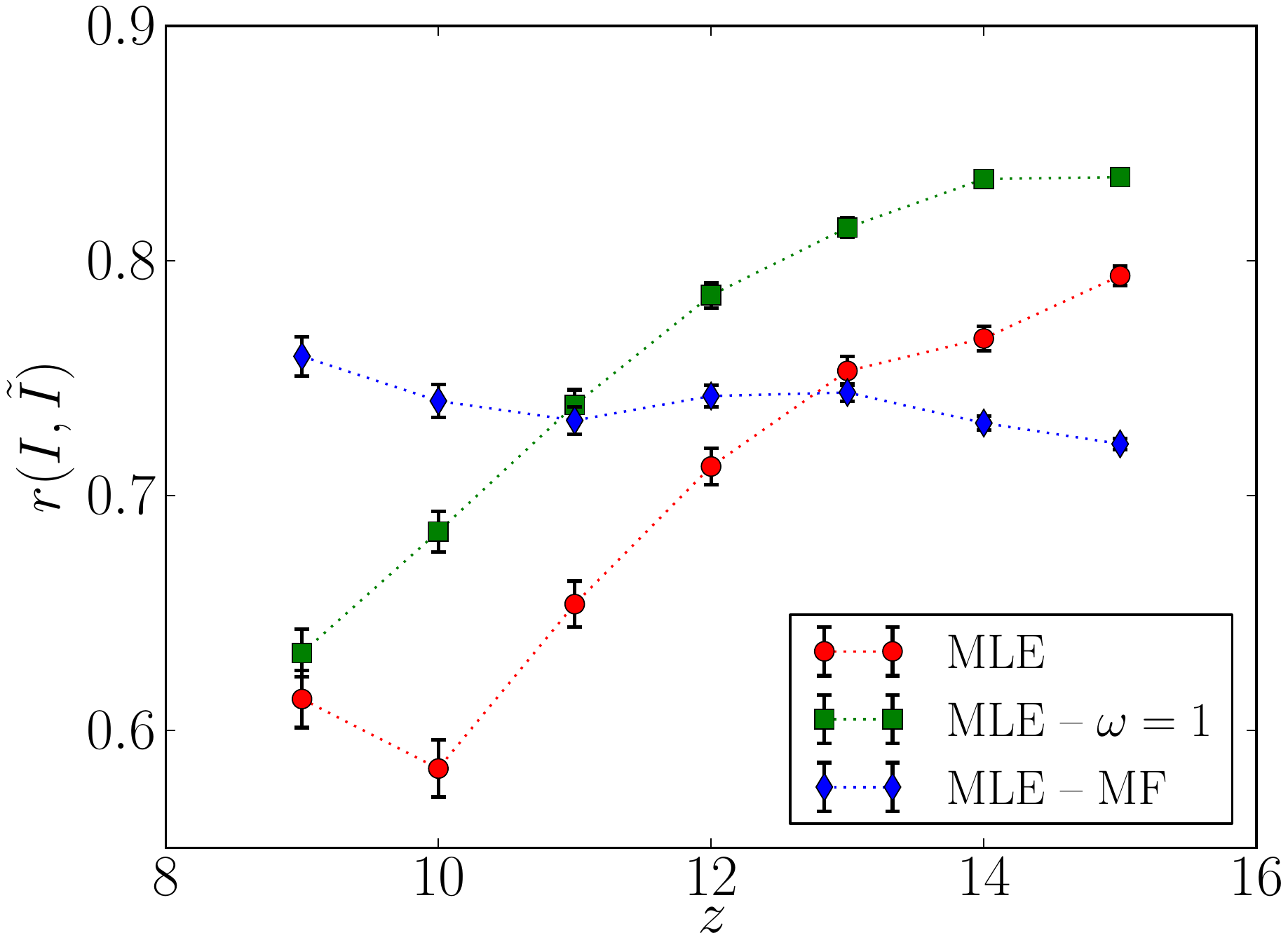}
\caption{Performance of estimation methods vs mean degree of artificial networks $z$.}
\label{fig:comparison}
\end{figure}

We proceed now to a direct comparison of the discussed estimation methods. As can be seen in Tab.~\ref{tab:evaluation}, all methods produce log-likelihood of a comparable magnitude but their parameter estimates (only $\tilde A$ and $\tilde\varTheta$ are shown here) differ. Notably, while $\tilde\varTheta$ is close to the true value, the error of $\tilde A$ is substantial. One can say that MLE tends to overestimate the tendency to random connections (as $A$ grows, the influence of preferential attachment vanishes) in this artificial system. We found no signs of this error vanishing with the data size which can be probably attributed to the network growth which constantly injects new nodes with zero degree in the system. The highest correlation between $I$ and $\tilde I$ is achieved with penalized MLE estimation when $\omega=1$. Performance of the methods is further illustrated in Fig.~\ref{fig:comparison} as a function of $z$. As expected, $r(I, \tilde I)$ increases with $z$ for both exact MLE methods and penalized estimations always outperform the unpenalized ones. The behavior is different for results obtained with the mean-field MLE whose quality slowly deteriorates as $z$ increases. Obviously, further improvements are necessary to make this otherwise promising method applicable in practice.

\begin{table*}
\centering
\begin{ruledtabular}
\begin{tabular}{rrrrrrrrrr}
model $M$ & $P_i(t)$ & $k_M$ & $\ln\mathcal{L}/E$ & $AICc(M)$ & $w_M$ & $\tilde A$ & $\tilde R_{\infty}$ & $\tilde\varTheta$ & $\tilde\beta$\\
\hline
 RAND &                $1$ &   $0$ & $-5.805$ & $285,364$ &  $0$ &    --- &      --- &   --- & ---\\
   PA &         $k_i(t)+A$ &   $1$ & $-5.767$ & $283,519$ &  $0$ &  $137$ &      --- &   --- & ---\\
 PA-H &       $k_i(t)+A_i$ &   $N$ & $-4.641$ & $229,931$ &  $0$ &    --- &      --- &   --- & ---\\
PA-HD &    $k_i(t)+A_i(t)$ & $N+2$ & $-4.111$ & $203,872$ &  $0$ &    --- &      --- & $1.9$ & $0.42$\\
 PA-R &    $(k_i(t)+A)R_i$ & $N+1$ & $-4.641$ & $229,905$ &  $0$ & $1108$ &      --- &   --- & ---\\
PA-RD & $(k_i(t)+A)R_i(t)$ & $N+4$ & $-4.043$ & $200,536$ &  $1$ &   $60$ & $0.0088$ & $3.2$ & $0.64$\\
\end{tabular}
\end{ruledtabular}
\caption{Maximum likelihood estimates and model selection results for the Econophysics Forum data. Model weights $w_M$ defined by Eq.~(\ref{w_M}) show that there is overwhelming evidence in favor of the model with decaying relevance (PA-RD).}
\label{tab:results}
\end{table*}

\section{Analysis of real data}
\label{sec:real_data}
To illustrate the potential of high-dimensional statistical analysis of network data, we finally apply it to real data and compare the level of support which it gives to various network models. The analyzed data originates from the interdisciplinary physics community web site \emph{Econophysics Forum} (see \url{www.unifr.ch/econophysics}) which is run by the research group of Yi-Cheng Zhang at the University of Fribourg since 1998. We parsed server web log files collected from 6th July 2010 until 31st March 2013 (a time span of 1000 days). Activities of web bots and other automated access were removed from the data. While web logs contain all user actions on the web site, we kept only entries corresponding to downloads of papers posted on the Econophysics Forum. The corresponding user-paper bipartite network consists of 844 paper-nodes and their 24,581 links~\cite{data}. As expected, the degree distribution of paper-nodes is broad (the maximal degree of 741 is much greater than the average degree of 29), making this data a good candidate for being explained by preferential attachment or related models.

We use this data to evaluate two low-dimensional and four high-dimensional models. The low-dimensional models are: random attachment to an existing node (RAND) and the standard preferential attachment (PA). The high-dimensional models are: preferential attachment with heterogeneous (node-dependent) additive term (PA-H), preferential attachment with heterogeneous and decaying additive term (PA-HD) which has been introduced  in~\cite{Eom2011}, preferential attachment with constant relevance (PA-R; such constant relevance is usually referred as fitness in past works~\cite{Bianconi01}), and finally preferential attachment with relevance decay (PA-RD). The functional form of the probability of a new link attaching to an existing node at time $t$, $P_i(t)$, is shown in Tab.~\ref{tab:results} for each model. The form of $A_i(t)$ suggested for PA-HD in~\cite{Eom2011} is generalized to $A_i(t) = I_i(t) \exp[-((t-\tau_i)/\varTheta)^{\beta}]$ which in our case performs better than the original form without $\beta$. Note that~\cite{Huberman13} reports a similar behavior in the popularity growth of stories in \url{digg.com}. For simplicity, we assume a similar form of $R_i(t)$ in PA-RD, $R_i(t) = I_i(t) \exp[-((t-\tau_i)/\varTheta)^{\beta}] + R_{\infty}$, which in fact roughly corresponds to the empirical relevance decay results presented in~\cite{Medo2011}. A non-vanishing absolute term $R_{\infty}$ is needed here to allow for links occasionally attaching to old nodes. The log-normal decay form reported in \cite{Radicchi08,Wang2013} does not yield better fit in our case, perhaps as a result of immediate response of the Econophysics Forum users which makes the increasing relevance phase provided by log-normal curves unfitting. For PA-RD, we report results obtained with the penalization term ($\omega=1$) which, however, differ little from the results obtained with $\omega=0$.

To maximize the likelihood functions we use the iterative extrapolating approach described in Sec.~\ref{sec:MLE}. This procedure is run ten times with independent random initial configurations; the best result obtained with each method is reported in Table~\ref{tab:results}. In addition, the table shows also the number of model parameters $k_M$ and the corrected Akaike information criterion
\begin{equation}
AICc(M) = -2\ln\big(\!\max\mathcal{L}(\mathcal{D}\vert M)\big)+\frac{2k_ME}{E-k_M-1}
\end{equation}
where the maximum is taken over the whole parameter space of model $M$. $AICc(M)$ measures how well model $M$ fits the data and corrects for a finite sample size~\cite{Burnham2002}. It can be used to construct model weights~\cite{Claeksens2008} in the form
\begin{equation}
\label{w_M}
w_M\sim\exp\big[(\min_{M'} AICc(M')-AICc(M))/2\big]
\end{equation}
where the proportionality factor is obtained by requiring the sum of all model weights to equal one. Finally, we report the values of global model parameters that maximize data likelihood for each model.

Our comparison of models contains several notable outcomes. Firstly, both low-dimensional models are clearly insufficient to explain the data. In fact, preferential attachment yields only marginally better fit than random attachment. Secondly, high-dimensional models without time decay perform significantly worse than their counterparts with time decay. This is not surprising because we fit the models to an information network where, as argued in~\cite{Medo2011}, aging of nodes is of prime importance. Thirdly, while the log-likelihood values obtained with PA-HD and PA-RD are both substantially better than those obtained for other models, the difference between them is big enough for the Akaike information criterion to assign an overwhelming weight to PA-RD. (The resulting weight of PA-HD, which has been truncated to zero in Table~\ref{tab:results}, is around $10^{-724}$.)

For PA-RD, the effective lifetime corresponding to the obtained relevance decay parameters is
\begin{equation}
\avg{t} := \frac{\int_0^{\infty} t R(t)\,\dd t}{\int_0^{\infty} R(t)\,\dd t} =
\frac{\tilde\varTheta\,\Gamma(2/\tilde\beta)}{\tilde\beta\,\Gamma(1+1/\tilde\beta)}\approx 8\,\mathrm{days}
\end{equation}
where we neglect $R_{\infty}$ which is small, yet it formally causes the above-written expression to diverge. This lifetime well agrees with the fact that papers typically spend one week on the front page of the Econophysics Forum. The value of the additive term $\tilde A \approx 60$ is relatively high in comparison with the average node degree of $29$ which suggests that in the studied dataset, the influence of preferential attachment (\emph{i.e.}, attachment probability proportional to node degree) is relatively weak. An alternative explanation is that our assumed relevance decay function $R(t)$ disagrees with the data and thus an increased proportion of ``random'' connections is necessary to model the data. A more detailed analysis is necessary to establish what is the real reason behind this apparent randomness.

\begin{figure}
\centering
\includegraphics[scale = 0.4]{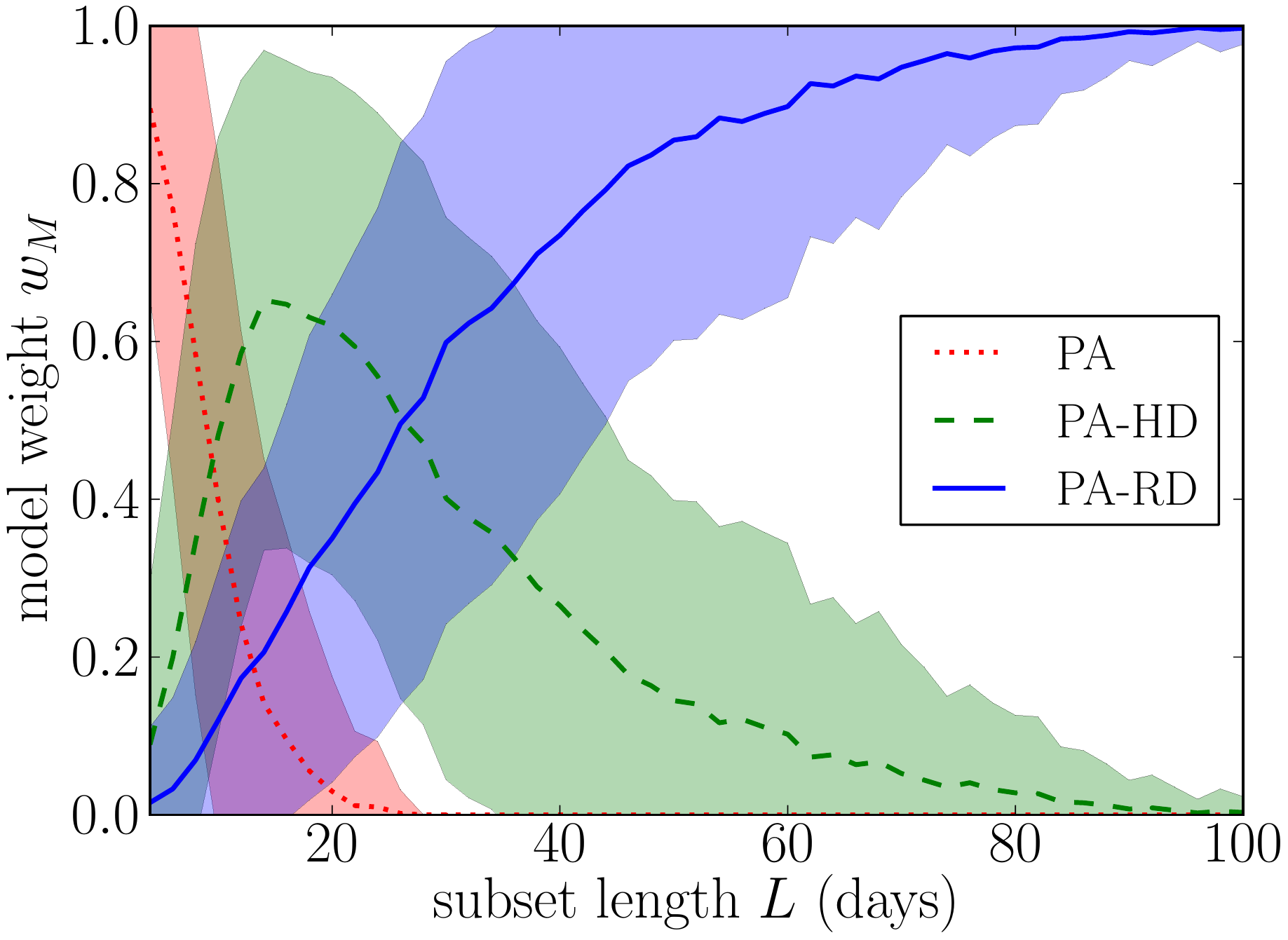}
\caption{Model weights as a function of subset length for PA, PA-HD, and PA-RD. Lines show the mean weights and shaded areas mark their standard deviation based on 1000 random subsets drawn for each subset length.}
\label{fig:weights}
\end{figure}

Since likelihood computation is costly and during its maximization in numerous variables it needs to be carried out many times, obtaining the results presented in Table~\ref{tab:results} on a standard desktop computer takes several hours. It is thus natural to ask whether significant evidence in favor of one of the models cannot be obtained by analyzing subsets of the data which would save considerable computational time. To this end, we evaluated weights of three representative models (PA, PA-HD, and PA-RD) on data subsets corresponding to time spans (which we refer to as subset lengths, $L$) ranging from 4 to 100 days; the starting. We generated many subsets for each $L$ by choosing their starting day at random. Results shown in Fig.~\ref{fig:weights} demonstrate that while particularly short subsets favor the low-dimensional PA model, the situation quickly changes and this model is virtually eliminated as soon as $L\gtrsim30$. Two high-dimensional models, which enjoy comparable support until $L\lesssim30$, are clearly distinguished at $L=60$ and above. Meanwhile, evaluation of multiple small-scale subsets is fast: the computational time required for one likelihood maximization of PA-RD drops from 10 minutes for the whole 1000-day data spanning to 2 seconds for a 100-day subset. We can conclude that this approach allows us to efficiently discriminate between models even when no particularly efficient approach to likelihood maximization is available.

\section{Discussion}
\label{sec:conclusions}
We studied the use of maximum likelihood estimation in analysis of high-dimensional models of growing networks. Artificially created networks with preferential attachment and decaying relevance~\cite{Medo2011}  were used to show that a near-flat likelihood landscape makes the standard likelihood maximization rather unreliable and sensitive to the initial choice of model parameters. Introducing a penalization term effectively modifies the landscape and helps to avoid ``wrong'' solutions. The resulting MLE-based scheme outperforms the standard likelihood maximization for a wide range of model networks. On the other hand, both original and modified MLE overestimate the additive parameter $A$ which is crucial in the early stage of a node's degree growth. How to improve on that remains an open question.

We then tested the previously developed methods on real data where both preferential attachment and relevance decay are expected to play a role. In this part, the focus is on comparing various competing network models that may be used to explain the data. We show that the data shows overwhelming evidence in favor of one of the models and that sufficiently strong evidence can be achieved by studying small subsets of the data. Model evaluation by such subset sampling is of particular importance to large-scale datasets where straightforward likelihood maximization is prohibitively time-consuming.

Up to now, models of complex networks have been appraised mostly by comparing aggregate characteristics of the produced networks (degree distribution or clustering coefficient, for example) with features seen in real data. The caveat of this approach is that many network characteristics are computed on static network snapshots and are thus of little use for the measurement of growing networks. Empirical node relevance~\cite{Medo2011} is designed especially for growing networks but more metrics, targeted at specific situations and questions, are needed.

Despite potential improvements in this direction, to gain \emph{real} evaluative and discriminative power over network models, robust statistical methods such as maximum likelihood estimation need to be relied on. We have made a step in this direction which, hopefully, will contribute to consolidating and further developing the field of network models. Open issues include estimates of parameter uncertainty in the case of real data by bootstrap methods~\cite{Davison97,Shalizi10}, identification of situations where maximum likelihood estimates converge to true parameter values (including model misspecification as in~\cite{Owhadi2013} which is of particular importance to parameter estimates in complex systems), and improvements of the mean-field likelihood estimation which was introduced in Sec.~\ref{sec:MF-MLE}. It needs to be stressed that the potential impact of parameter estimation far exceeds the academic problem of model validation: Model parameters, once known, can be directly useful in practice. In the case of preferential attachment with relevance decay, for example, the overall rate of relevance/interest decay is closely connected to the most successful strategy in the competition for attention~\cite{Huberman13}. On the other hand, the initial, current, or total relevance values of individual items can be used to detect which items deserve to be examined more closely.

\begin{acknowledgments}
This work was supported by the EU FET-Open Grant No.~231200 (project QLectives) and by the Swiss National Science Foundation Grant No.~200020-143272.
\end{acknowledgments}

\appendix

\section{On the shape of the likelihood function}
\label{sec:maxima}
Greedy sequential optimization is possible because the likelihood function in our case does not have a large number of disparate local minima. We explain this fact for the PA-RD model which is parametrized by the initial node relevances $I_i$ and global parameters $A$ and $\varTheta$. \cite{Medo2011} shows that the expected final degree of node $i$ grows with $I_i$ which implies that $\mathcal{L}(\mathcal{D}\vert \vek{I}, A, \varTheta)$ has a unique maximum in $I_i$ when all other parameters are fixed. Likelihood of the data thus has a unique maximum in the space of all initial relevance values. Similar behavior can be observed for $A$. When $A\to\infty$, likelihood of the artificial data is small because the model simplifies to random attachment which is obviously at odds with the data. As $A$ decreases, the likelihood grows but it eventually saturates and decreases when $A$ becomes so small that new nodes cannot attract their first links. The case is different for $\varTheta$. Its extremely small values can be easily refuted by the data as they would imply links always arriving at the latest node. On the other hand, large $\varTheta$ can be accommodated by an appropriate choice of the initial relevance values which is demonstrated by Fig.~\ref{fig:profile}. To prevent the sequential updating of parameters from converging to a wrong solution, one can for example add a suitable penalization term as we do in \eref{LLpenalized}.

\end{document}